\documentclass{article}

\usepackage[margin=0.75in]{geometry}

\usepackage{amsmath}
\usepackage{graphicx}
\usepackage{multirow}
\usepackage{dsfont}

\usepackage{natbib}
\setcitestyle{authoryear, open={((},close={)}}

\usepackage{setspace}
\usepackage{amsthm}
\usepackage{thmtools}
\declaretheoremstyle[
  spaceabove=10pt,
  spacebelow=10pt,
  headfont=\bfseries\itshape,
  notefont=\bfseries\itshape,
  headpunct={},
  postheadspace = \newline,
]{mystyle} 

\declaretheorem[name={Definition},style=mystyle]{definition}
\declaretheorem[name={Example},style=mystyle, sharenumber=definition]{example}

\usepackage{fancyhdr}
\title{Conceptualizing experimental controls using the potential outcomes framework}
\author{Kristen B. Hunter\footnote{Department of Statistics, Harvard University}, Kristen Koenig\footnote{Department of Organismic and Evolutionary Biology, Harvard University}, Marie-Ab\`{e}le Bind\footnote{Biostatistics Center, Massachusetts General Hospital}}
\date{\today}

\begin{document}

\maketitle

\section*{Abstract}
The goal of a well-controlled study is to remove unwanted variation when estimating the causal effect of the intervention of interest.
Experiments conducted in the basic sciences frequently achieve this goal using experimental controls, such as ``negative'' and ``positive'' controls, which are measurements designed to detect systematic sources of such unwanted variation.
Here, we introduce clear, mathematically precise definitions of experimental controls using potential outcomes.
Our definitions provide a unifying statistical framework for fundamental concepts of experimental design from the biological and other basic sciences.
These controls are defined in terms of whether assumptions are being made about a specific treatment level, outcome, or contrast between outcomes.
We discuss experimental controls as tools for researchers to wield in designing experiments and detecting potential design flaws, including using controls to diagnose unintended factors that influence the outcome of interest, assess measurement error, and identify important subpopulations.
We believe that experimental controls are powerful tools for reproducible research that are possibly underutilized by statisticians, epidemiologists, and social science researchers.

\textbf{Keywords:} negative control; positive control; causal inference; experimental design.


\section{Introduction}

\subsection{Motivation}

In 2013, a meta-analysis of the relationship between weight and mortality found that being overweight, but not obese, was associated with a lower mortality than being normal or underweight \citep{Flegal2013}.
These results contradicted medical consensus that being overweight leads to poor health outcomes \citep{GlobalBMI2016}.
Subsequent investigations into the analysis discovered that the population of normal and underweight people in the study contained a high proportion of people who were smokers, elderly, or chronically ill, leading to misleadingly high mortality rates in those groups \citep{Miller2013}.
This meta-analysis study is an example of how a study can have a fundamental flaw, even when researchers have the best intentions.
The researchers were interested in isolating the effect of weight on mortality, but failed to design a study that compensated for other factors that also influenced mortality.
The goal of this work is to discuss one tool to assist researchers in designing high-quality studies: experimental controls.

The goal of a well-controlled study is to create a context in which the intervention of interest is the sole causal mechanism influencing the outcome of interest.
In this work, we discuss using experimental controls to design robust scientific studies.
The most common use of the term control in a statistical study is to refer to a particular treatment group which receives a control treatment level, such as a placebo or no active treatment \citep{Boring1954}.
Here, we use the term control to refer to an experimental control, which is an intentional set of experimental conditions, such as treatment levels and outcomes, that encode a scientific assumption about a study.

The goal of using an experimental control is to attempt to detect or remove systematic sources of variation due to extraneous factors \citep{Kirk1982}.
By factor, we mean an experimental condition or setting.
The controls we introduce are defined based on a combination of outcomes and specific treatment levels, rather than being singularly defined based on a treatment level, as is the case for the most common statistical interpretation of the term control.
Although the controls that we discuss encompass many different concepts, they all share the property that they describe phenomena that are already well-understood through previous experiments or studies before the study is conducted.
We focus our discussion on controls in randomized experiments, but the concepts translate easily to other study types, including observational studies such as the one mentioned above.

In this work, our main contribution is to outline clear, mathematically precise definitions of different types of controls using the potential outcomes framework \citep{Neyman1923, Rubin1974}.
The diverse set of disciplines that have explored experimental controls has also resulted in variation in the terminology, definitions, and uses of controls across fields, which can result in a lack of clarity and understanding.
We give researchers an overview of controls, and classify different types of controls in one consistent framework.
The definitions we introduce involve mathematical statements, but still can be easily translated into practical use.
These definitions are motivated by the uses of controls in basic laboratory science, particularly biology.
Our second contribution is to discuss some practical uses of controls as diagnostic and design tools in experiments.
We illustrate how controls can be used to detect possible experimental flaws, and how they can help design a more robust experiment by helping to identify and eliminate both systematic and random errors.
Specifically, we discuss using controls to diagnose unwanted factors that influence the outcome, assess measurement error, determine optimal timing of an intervention, identify responders, and identify compliers.

Controls have previously been discussed and used in useful and meaningful ways in causal inference settings.
However, defining them within the potential outcomes framework is novel.
Statisticians have mostly used controls in the context of directed acyclic graphs (DAGs); see \citet{Lipsitch10} for an introduction to controls using DAGs, and \citet{Shi2020} for a recent review of negative control methods in epidemiology.
Research in this area has produced robust statistical methods, in particular in improving the validity of the statistical analyses of observational studies. 
Although DAGs are often a useful analytical framework, we believe defining controls within the potential outcomes framework provides complementary value.
Many researchers are not as familiar with the interpretation of DAGs.
Additionally, the potential outcomes framework allows for definitions that are simple, easy to interpret, and mathematically precise.
We hope that by clearly encoding assumptions, this work can potentially lead to easier communication between practitioners and analysts.

Returning to our introductory example, how could controls have prevented the error at hand?
We follow a hypothetical example.
One type of control we will introduce below is a null contrast-control, which is an effect that we assume to be zero given the treatment of interest.
A causal effect is a comparison of an outcome given different levels of treatment.
Our control assumption is that weight has no effect on the rate of deaths due to motor vehicle accidents.
Previous research has found no link between obesity and mortality after trauma \citep{Alban2006}, although a null result can never be definitively proven.
Thus, we test the null hypothesis that the rate of death by motor vehicle accident is same in the overweight group as in the normal/underweight group.
We find that the rate of deaths by motor vehicle accidents is much lower in the normal/underweight group than the overweight group, and after further investigation discover this difference is induced by lower rates of motor vehicle deaths among elderly people, who are more heavily represented in the normal/underweight group.
Ideally, researchers would have conducted this test during the design stage of their study, leading them to discover the imbalance in their study populations and re-design the study.

Unwanted variation can often be accounted for in two ways during the design stage of a study: increasing the size or changing the composition of the sample, or implementing the use of controls.
Two of the main sources of variation considered by biologists are biological and technical variation, which are explored using biological and technical replicates.
Including biological replicates means repeating the study across different samples to account for ``the intrinsic biological variability of a system'', while technical replicates refers to the repeating of the study across different experimental conditions to account for the variation independent of biological variability [Chapter 29]\citep{Glass14}.
In addition to increasing statistical power, replicates can potentially help reduce bias if they are designed to ensure a sufficiently large and diverse set of data is included.
Controls and replicates are often complementary tools: for example, controls can help detect unwanted variation, which then can be accounted for using replicates.
In the above example of obesity, a larger study size with the same selection criteria would not have fixed the problem.
However, a control could have helped detect the flawed design, which would have informed how to target a set of biological replicates to include a more representative population.
Although replicates are an important consideration in experimental design, in this work we focus on controls, which detect or compensate for unwanted variation without changing the size or composition of the sample.

This work is intended to anchor mathematical definitions of controls, so researchers are aware of and can easily apply a broader set of controls than is typically used in causal inference.
Our explanation of these concepts comes from the basic sciences, specifically biology, where experimental controls are much more commonplace.
We advocate for statisticians include more controls in their arsenal of tools.
We believe that the extra information provided by measuring controls can lead to a deeper understanding of causal mechanisms, and more robust and reproducible conclusions.
In particular, experimental controls provide avenues for detecting and preventing weak study designs.
A poorly-controlled study may result in less precise results due to extra unwanted variation, and in the worst case can result in misleading conclusions.

\subsection{Motivating example and review of potential outcomes framework}

As a motivating example, we consider the question:~what is the short-term effect of drinking caffeinated coffee compared to drinking decaffeinated coffee on a person's change in blood pressure from a baseline level?
For this motivating example, we follow the helpful hypothetical experiment discussed in detail in \citet{Glass14} (Section Three: The Experiment).
We use this hypothetical example throughout the paper to illustrate different types of controls that can be used in experimental design, capitalizing on the potential outcomes framework.
This section provides a brief introduction to the potential outcomes framework through our motivating example.

The random variable indicating treatment assignment for unit $i$ is $W_i$.
For the running example, our units are people, so we will use the terms units and people interchangeably.
If $W_i = w_{at}$, unit $i$ has been assigned to the active treatment, and if $W_i = w_{ct}$, unit $i$ has been assigned to the control treatment.
Here, the active and control treatments are drinking caffeinated coffee and drinking decaffeinated coffee, respectively.

In an experiment, each unit can be conceptualized as having different potential outcomes given each possible treatment assignment.
This framework is known as the Neyman-Rubin Causal Model \citep{Neyman1923, Holland86}.
The primary outcome of interest ($p$), which we denote $Y_i^{p}$, is the change in blood pressure measured before and after an intervention.
If unit $i$ were to be assigned to the active treatment $w_{at}$, we denote the potential outcome given treatment $w_{at}$ to be $Y_i^{p}(w_{at})$.
Similarly, the potential outcome given the control treatment $w_{ct}$  is $Y_i^{p}(w_{ct})$.
We assume SUTVA, which refers to the stable unit treatment value assumption \citep{Rubin1980}.
We expand the framework to allow for additional outcomes and treatment levels, which will be defined in Section~\ref{sec:def}: for now, let $Y_i^{k}(w_j)$ denote the potential outcome $k$ for unit $i$ had unit $i$ been exposed to treatment level $w_j$.

Our primary estimand of interest, $\tau^p$, is the
average difference between the active treatment level and control treatment level for the outcome of interest.
Specifying a causal effect, which we will refer to as an effect for simplicity, requires specifying an outcome and a comparison between two treatment levels.
The unit-level causal effect of primary interest is $$\tau_{i}^{p} =Y_{i}^{p}(w_{at})-Y_{i}^{p}(w_{ct}).$$
Given $N$ units, then the causal effect of primary interest is $$\tau^{p} = \frac{1}{N}\sum_{i=1}^N \tau_{i}^{p}=\overline{Y}^{p}(w_{at})-\overline{Y}^{p}(w_{ct}).$$
Here, this is the average difference in blood pressure between the active treatment of drinking caffeinated coffee and the control treatment of drinking decaffeinated coffee across units.
The analyst may also estimate causal estimands of secondary interest, which can be used for descriptive purposes or for validation; we will discuss some potential secondary estimands later.

The fundamental problem of causal inference is that we can only observe one of these potential outcomes per unit, because we cannot simultaneously assign more than one treatment \citep{Rubin2008b}.
Thus, for each unit $i$ and each outcome $Y_i^{k}$, only one of the $J$ potential outcomes can be observed.
We finally define the observed outcomes, $Y_i^{k,obs} = \sum_{j=1}^{J} \mathds{1}(W_j = w_j) Y_i^k(w_j)$, where $\mathds{1}(x)$ is an indicator variable for condition $x$.
The average observed outcome given treatment $w_j$ is $\overline{Y}_{w_j}^{k,obs} = {N_{w_j}}^{-1} \sum_{i:W_i = w_j} Y_i^{k,obs}$, where $N_{w_j}$ is the number of units assigned to treatment $w_j$.

\section{Definitions}
\label{sec:def}

\subsection{Introduction}

We introduce controls defined on two different axes: first, whether they are null controls or non-null controls; and second, whether they are treatment, outcome, or contrast controls (Table~\ref{tab:class}).
The first category pertains to whether the results are known to be zero or nonzero, while the second defines what aspect of the experiment is being controlled.
For all introduced controls, we provide an example using the hypothetical caffeine experiment.

\begin{table}
\centering
\begin{tabular}{l r r}
& Null                                              & Non-null \\
Treatment
& null treatment-control                            & non-null treatment-control \\
& $Y_i^{p}(w_{nt}) = 0$                             & $Y_i^{p}(w_{nnt}) \neq 0$ \\
Outcome
& null outcome-control                              & non-null outcome-control \\
& $Y_i^{no}(w_{at}) = 0$                            & $Y_i^{nno}(w_{at}) \neq 0$ \\
Contrast
& null contrast-control                             & non-null contrast-control \\
& $\tau_i^{nc}= 0$                                  & $\tau_i^{nnc}\neq 0$\\
\end{tabular}
\caption{Classifying controls}
\label{tab:class}
\end{table}

\subsection{Treatment-controls}

\begin{definition}[null treatment-control]
\label{def:ztc}
A null treatment-control is a treatment level $w_{nt}$ for which the primary outcome $Y_i^{p}$ is zero for unit $i$.
\[Y_i^{p}(w_{nt}) = 0.\]
\end{definition}

\begin{example}
In the caffeine example, no intervention (i.e. remaining quietly seated) is a null treatment-control on change in blood pressure.
If no intervention occurs, we are making the assumption that the true underlying blood pressure of a person does not change.
\end{example}

After introducing the first type of control, we now discuss some important aspects of controls and the corresponding definitions.
First, we are defining controls based on the true, latent potential outcomes.
However, in reality we may observe the outcome with measurement error.
In the caffeine example, we may make the assumption that a person's underlying blood pressure would not change with no intervention, but we could easily observe a change due to natural fluctuations and measurement error.
Causal inference in a setting in which the outcome has measurement error is not the focus of this work; see for example \citet{Shu2019} for more discussion.
Second, we define the control at the unit level.
By defining controls at the unit level, we do not require the assumption to hold for all units.
We will further discuss these two topics in Section~\ref{sec:diag_error}.

In this case, we are defining a null treatment-control as a case when the primary outcome is zero, but we could also consider broader generalizations.
Conceptually, the definition of a null treatment-control is that it is a treatment resulting in a null outcome.
In a different context, a null outcome may not be zero; for example, if the outcome is a ratio of blood pressure measurements, the null treatment-control assumption would be that the outcome is one rather than zero.

In using the terminology of a `null' treatment-control, we depart from the more standard terminology of calling this type of control a negative control.
In standard terminology, the control introduced in the next section, a non-null treatment-control, would be a type of positive control.
Though the terms negative and positive control are more traditionally used by scientists and the statistical community, we are using new terms to emphasize that the mathematical definitions do not correspond to positive or negative quantities.
As discussed above, a null contrast-control could be zero or any quantity that corresponds to a `null' value; similarly, a non-null contrast control can have a positive or negative value.

\begin{definition}[non-null treatment-control]
\label{def:nnntc}
A non-null treatment-control is a treatment level $w_{nnt}$ for which the outcome of interest $Y_i^{p}$ is nonzero for unit $i$.

\[Y_i^{p}(w_{nnt}) \neq 0 \]
\end{definition}

\begin{example}
Taking a hypertensive drug (e.g., a drug that raises blood pressure) is a non-null treatment-control on change in blood pressure.
If a patient is given a hypertensive drug that is a priori known to be effective for that patient, their underlying blood pressure will increase.
\end{example}


\subsection{Control treatment levels}

Control treatment levels are essential to make a causal estimand well-defined, and are a distinct concept from treatment-controls.
A control treatment level is an alternative treatment level that we would like to compare to the active treatment level.
As stated in \citet{Rubin2008b}, ``a causal effect is, by definition, a comparison of treatment and control potential outcomes on a common set of units.''
Thus, the control treatment level is essential to define the estimand: in our example, what is the effect of drinking caffeinated coffee on blood pressure, ``compared to what?'' \citep{Feller16}.
The choice of control treatment level isolates the particular effect of interest.
For example, consider the following possible control treatment levels and their corresponding estimands in Table~\ref{tab:control_treatments}, which continue to follow \citet[Chapter 23]{Glass14}.

\begin{table}
\centering
\begin{tabular}{lll}
Control treatment                   & Estimand: $Y(w_{at}) - Y(w_{ct})$ \\ & \\
$w_{\text{null}}$: No intervention  & Effect of all components of coffee \\
$w_{ct1}$: Water                    & Effect of non-fluid component of coffee \\
$w_{ct2}$: Caffeinated water        & Effect of non-caffeine component of coffee\\
$w_{ct3}$: Decaffeinated coffee     & Effect of caffeine component of coffee
\end{tabular}
\caption{Control treatments}
\label{tab:control_treatments}
\end{table}

The choice of control treatment level is motivated by the question of interest.
We have chosen $w_{ct3}$, decaffeinated coffee, as a control treatment level, which gives an estimate of just the isolated effect of the caffeine in coffee.
Alternatively, choosing $w_{ct1}$, water, expands our estimand to include the effect of all the non-fluid components of coffee, such as any other compounds in coffee in addition to caffeine.

Treatment-controls can be used as control treatment levels, but usually are not chosen because they are often not as useful as a point of comparison.
For example, comparing caffeinated coffee to no intervention gives an estimate of all components of coffee, and would not isolate caffeine alone.
Drinking fluids can actually temporarily increase blood pressure in some people \citep{Ma2019}, so by not comparing coffee to another fluid, we estimate the combination of caffeine, other components in coffee, and the fluid effect.

A common choice of control treatment level is a placebo treatment.
\citet{Rubin2019} formally defines a causal placebo effect as
$$Y(W_i = w_{ct}, \Pi = \pi) - Y(W_i = w_{ct}, \Pi = 0),$$
where $W_i$ is the primary treatment, $\Pi$ is the patient's probability of being assigned to the active treatment, which is known to the patient, and $\pi$ is a particular realized value of the probability the patient is assigned to the active treatment, which may be unknown to the patient.
The placebo effect is the difference in outcomes between a subject knowing that they may receive the placebo with probability $1 - \pi$, and knowing that they definitely receive the placebo.
The second treatment, the patient knowing that they receive the placebo, is a null treatment-control.
Thus, in order to have the ability to estimate this causal placebo effect, the outcome given the null treatment-control must be measured.
However, clinical trials do not usually incorporate null treatment-controls: there is usually no study arm in which the patients know they are receiving the placebo, and thus most studies are unable to estimate the placebo effect.
\citet{Rubin2019} encourages researchers to include an unblinded arm in RCTs focusing on dosage recommendations, which would allow the additional outcome given the null treatment-control to be measured.
Comparing a new drug to a placebo is important in demonstrating its efficacy, but does not represent the reality of clinical practice, where the choice is between the drug and no intervention.
By estimating the placebo effect using a null treatment-control, we can disentangle the placebo component from and the medical effects of a drug \citep{Mozer18}.
Isolating these two effects can lead to a better treatment regime, and lessen the risk that a patient is prescribed a higher dose than is necessary.


\subsection{Outcome-controls}
\label{sec:outcome-control}

\begin{definition}[null outcome-control]
\label{def:zoc}
A null outcome-control is an outcome $Y_i^{no}$ that is zero for unit $i$ given the active treatment level $w_{at}$.
\[ Y_i^{no}(w_{at}) = 0 \]
\end{definition}

We emphasize that $Y_i^{no}$ is a distinct outcome from the primary outcome of interest $Y_i^{p}$.

\begin{example}
Change in body flexibility is a null outcome-control for drinking caffeinated coffee.
The current scientific understanding is that body flexibility does not change after drinking coffee.
\end{example}

\begin{definition}[non-null outcome-control]
\label{def:noc}
A non-null outcome-control is an outcome $Y_i^{nno}$ that is nonzero for unit $i$ given the active treatment level $w_{at}$.
\[ Y_i^{nno}(w_{at}) \neq 0 \]
\end{definition}

\begin{example}
Change in electrolyte concentration is a non-null outcome-control for drinking caffeinated coffee.
Electrolyte concentration is known to change after drinking coffee.
\end{example}


\subsection{Contrast controls}

\begin{definition}[null contrast-control]
\label{def:zec}
A null contrast-control is an effect $\tau_i^{nc}$ which is zero for unit $i$.
It is defined as a contrast of potential outcomes between an active treatment level $w_{atc}$ and a control treatment level $w_{ctc}$.
\[ \tau_i^{nc} = Y_i^{nc}(w_{atc}) - Y_i^{nc}(w_{ctc})= 0\]
\end{definition}

One important aspect of contrast-controls is they can never fully be observed directly: they must always be estimated, just as with any causal effect, because they are defined on a pair of outcomes, only one of which can be observed at a time.
We leave the definition purposefully broad in the choice of outcome, active treatment level, and control treatment level.

\begin{example}
The effect of caffeinated coffee compared to decaffeinated coffee on short-term change in body weight is a null contrast-control.
\end{example}

\begin{example}
The effect of caffeinated coffee compared to caffeinated water on change in blood pressure can also be a null contrast-control. This is a valid null contrast-control only if we believe from prior evidence that the only component of coffee that affects blood pressure is caffeine.
\end{example}

In the first example, we chose a secondary outcome, the primary active treatment level, and the primary control treatment level.
In the second example, we chose the primary outcome, the primary active treatment level, and a secondary control treatment level.

\begin{definition}[non-null contrast-control]
\label{def:nec}
A non-null contrast-control is an effect $\tau_i^{nnc}$ which is nonzero for unit $i$.
\[\tau_i^{nnc} = Y_i^{nnc}(w_{atc}) - Y_i^{nnc}(w_{ctc}) \neq 0\]
\end{definition}

\begin{example}
The effect of caffeinated coffee compared to decaffeinated coffee on change in reaction time is a non-null contrast control \citep{Santos14}.
\end{example}

Contrast-controls are related to outcome-controls in that they involve the choice of an outcome with specific properties, but they involve fundamentally different assumptions because outcome-controls are an absolute property, while contrast-controls are relative.
We restrict the following discussion to contrast-controls involving primary active treatment level and control treatment level.
For outcome-controls, only the active treatment is considered; we choose an outcome such that $Y_i^{nno}(w_{at}) \neq 0 $.
For contrast-controls, the difference between the active treatment and control treatments is considered; we are making an assumption about an effect, so choose an outcome such that the effect $Y_i^{nnc}(w_{at}) - Y_i^{nnc}(w_{ct}) \neq 0$.
Thus, in the caffeine example, outcome-controls are chosen based on any physical response to drinking coffee, which can include changes due to fluid, caffeine, or anything else in coffee.
Contrast-controls are chosen based on the response to caffeine alone.
To provide a concrete example illustrating the difference, we can use the outcome of electrolyte concentration for both non-null outcome-control and null contrast-control assumptions.
Electrolyte concentration is known to increase after drinking coffee, so as an outcome it is a non-null outcome-control.
On the other hand, the effect of caffeine on electrolyte concentration is a null contrast-control.
The influence of the coffee on electrolyte concentration is solely through the fact that someone is drinking a cup of fluid; based on current medical knowledge, caffeine does not play a role.
Thus, drinking caffeinated coffee or decaffeinated coffee results in the same change in electrolyte concentration, so it is a null contrast-control because the effect of caffeine on change in electrolyte concentration is zero.

\subsection{Discussion}

We introduce controls defined on two different axes: null controls and non-null controls, and treatment, outcome, and contrast controls.
Null controls are \textit{a priori} assumed to be zero; non-null controls are \textit{a priori} assumed to be not zero.
A treatment control is a treatment level that results in a known response for the outcome of interest.
An outcome control is an alternative outcome which has a known response to the active treatment level.
A contrast control is an alternative effect: when there is a known contrast between an active treatment level compared to a control treatment level on a particular outcome.

Some of our defined controls correspond to negative control outcomes, negative control exposures, positive control outcomes, and positive control exposures.
These terms have previously been defined and applied in causal inference literature \citep{Lipsitch10, Tchetgen13, Miao18}.
All of these terms are special cases of contrast-controls.
If we choose the outcome to be the primary outcome of interest, $Y^p$, but choose alternative active and control treatment levels, then we have a negative control exposure.
If we choose the outcome to be a secondary outcome, but choose the primary active and control treatment levels of interest $w_{at}$ and $w_{ct}$, then we have a negative control outcome.
We have chosen to introduce new terminology to make clear that this type of control is making an assumption about an effect, which is a contrast between outcomes given different treatment levels.
Based on our understanding, treatment-controls and outcome-controls do not directly translate into previously defined causal frameworks.

By choosing controls that follow our proposed definitions, the designer of the experiment is making assumptions based on prior knowledge.
If these assumptions fail, there are at least two plausible reasons: one is that the prior knowledge was incorrect; the second is that there was a flaw in the experiment.
Thus, controls should be carefully chosen by a practitioner to optimize the quality of the experiment.
Poorly chosen controls can either fail to detect flaws, or result in an experiment being rejected when in fact the prior knowledge was incorrect.
Some specific uses of controls to help ensure experimental quality is discussed in detail in Section~\ref{sec:design}.

If we assume that the underlying scientific assumption of a control is correct, then if a control assumption is observed to be violated, we have strong evidence that the experiment failed or has some defect.
In contrast, a control assumption being met cannot prove that the experiment was not flawed.
However, if we have an experiment with multiple well-chosen controls, we may deem the data to be of higher quality.
Many of the previous examples we have provided are contrived, and may seem unlikely to be useful in many settings.
When designing an experiment, we do not necessarily advocate that all types of controls must be used at all times.
Instead, the analyst must carefully consider the most likely points of error.
Thoughtfully-chosen controls can then help increase confidence that the experiment was not subject to these weaknesses.
If resources allow, we encourage experimenters to implement multiple types of controls, because some types are specific to detecting certain errors.
The more complementary measurements other than our main quantities of interest, the more likely we are to detect issues with our study.

\subsection{Controls in genetics studies}

Though we have focused on a hypothetical setting, we would like to emphasize that controls are already widely used in practice in both the design and analysis of experiments.
In particular, the use of experimental controls in genomics settings is a large and active research area.
We do not aim to fully review the large body of research in this area; instead, we briefly discuss some applications of controls in genomics, focusing on gene expression data, to illustrate the practical use of some of the controls we have defined.

Gene expression studies usually aim to determine the relationship between some disease status, in this setting the ``treatment'', and gene expression levels, which are the outcomes.
Gene expression data is known to be noisy and suffer from substantial technical variation, which is when samples have systematic patterns due to experimental conditions, such as which machine is used or at what time the sample is run, rather than underlying biology \citep{Lazor2012, Goh2017}.
Thus, researchers use different types of controls to help normalize data so that data sets from different machines, times, laboratories, or other settings can be more directly compared \citep{Gagnon12}.

To consider treatment-controls in genomics settings, we expand our definition of treatment to include any intervention that affects our outcome of measured gene expression levels, including interventions that occur in the laboratory setting rather than on the experimental subjects directly.
A spike-in control is a known quantity of RNA that is `spiked in' to the DNA microarray, which should result in a particular level of DNA expression \citep{Yang2006, Pine2016}.
Spike-ins are an example of technical controls, which target technical variation unrelated to biology.
Under our expanded definition of treatment, a spike-in control is a type of treatment-control.
A spike-in control can be considered a non-null treatment-control because it is a modification to the experiment that results in a known, non-null outcome of gene expression.
However, due to noise we usually observe a deviation from this expected level, which allows researchers to quantify the noise and normalize the rest of the expression data accordingly.

Outcome-controls and contrast-controls are also commonly used in genetics studies.
An outcome-control is a gene that has a known expression level.
For example, housekeeping genes regulate basic biological functioning and cellular processes, so they have expression levels that are relatively constant across people \citep{Lin2019, Eisenberg2013}.
However, we do have to be careful even with housekeeping genes; it is possible that systemic diseases like cancer could affect the expression of even these genes.
A contrast-control is a gene whose expression would be known to be affected or unaffected by the disease of interest, such as genes that are not differentially expressed in cancer and non-cancer patients.

\subsection{Related work}

Our work is in part inspired by compelling uses of controls in other frameworks.
\citet{Lipsitch10} proposed using negative controls in causal inference for the design of observational studies, focusing on epidemiological applications, and using negative controls is now increasingly common in epidemiological studies.
Following this work, \citet{Tchetgen13}, \citet{Richardson14}, \citet{Miao18}, and others have contributed statistical methods which use negative controls to help reduce confounding bias in observational studies, although such methods are not yet widely used in practice.
\citet{Vandenbroucke2016} also advocate for the use of negative controls in epidemiological studies to strengthen causal conclusions by ruling out alternative explanations.

\citet[Chapter 6]{Rosenbaum2002} provides a thoughtful discussion on using known effects to detect bias in observational studies.
One idea he discusses is testing the effect of the intervention of interest on an outcome known to be unaffected by that intervention, which can detect hidden bias or confounding.
This type of test is sometimes called a placebo test, and is sometimes applied in the fields of political science and economics.
However, as noted by \citet{Hartman2018}, there is no clear consensus for the definition of a placebo test, and different authors may apply the term differently.
Related concepts are refutability and falsification tests, which use empirical data to see if key assumptions do not hold \citep{Angrist1999}.
Similar to controls, falsification tests cannot prove that a key assumption is true, but they are useful checks to ensure that at least key assumptions are not provably false.
For example see \citet{Yang2014} and \citet{Keele19} for illustrations of falsification tests for instrumental variables.
Finally, negative controls have also been used to calibrate potentially noisy data.
\citet{Schuemie14} use negative controls to empirically calibrate p-value distributions.
\citet{Gagnon12} use negative controls to correct for unwanted variation in microarray data, and a series of later developments extends these methods to a wide variety of omics settings.

\citet{Rosenbaum2010} proposes the use of evidence factors, which are a series of statistically independent tests of the same null hypothesis.
The goal of an evidence factor is to use independent information to bolster the evidence of a particular conclusion by seeing if the conclusion holds under multiple different sets of assumptions, rather than being beholden to a particular set of possibly incorrect assumptions.
Here, we instead consider a series of different null hypotheses.
However, the goal is philosophically similar: to strengthen the reliability of a conclusion by considering different sets of assumptions and examining different pieces of the puzzle.

\section{Using controls in experimental design}
\label{sec:design}

\subsection{Background}

Having introduced a broad array of experimental controls, we now motivate why controls should be an important component of a researcher's toolbox.
We specifically discuss using controls as tools in the following cases: identifying unwanted factors, ensuring quality control when there is measurement error, determining optimal timing, identifying responders, and identifying compliers.
Following the rest of the paper, we use the hypothetical experiment of how caffeinated coffee influences short-term blood pressure compared to decaffeinated coffee.

Some uses of controls fall into the category of diagnostic tools, where the researcher uses a control to help detect potential experimental flaws.
If a proposed diagnostic test flags an issue, it speaks to a possible fundamental flaw in the experimental design, and thus the results from the experiment may not be considered valid in that they may not be successfully identifying the effect of interest. 
Ideally, a problematic diagnostic test would result in the researcher repeating the experiment with a new design to remove the potential flaw, although we recognize this is not feasible in many settings.
For example, in a small-scale biological experiment on fruit flies, it may be easy to repeat a given experiment many times, readjusting it based on the control variable diagnostics.
In a simple case, if a control diagnostic test points out a systematic measurement error, then the corresponding measurement tool can be recalibrated before the experiment is repeated.
However, in many fields, such as the social and biomedical sciences, studies are often difficult to replicate due to cost or feasibility.
Examples include large-scale observational studies and cost-intensive interventions, such as implementing a new education or economic program, or a new regulation.

In cases where a study is difficult to replicate, we advocate that, if feasible, researchers run a small pilot study in which they measure as many controls as possible.
During a pilot study, the full experiment is conducted on a small set of units, usually not the same ones that will be in the main experiment.
The cost of running such a pilot study and detecting flaws before they occur may well be worth the benefit of collecting data that is more trustworthy, rather than uncovering flaws after the fact.

Other uses of controls are more helpful in carefully choosing the design and parameters of a study.
For example, we can use controls to determine the best timing to measure the outcome.
In some cases, this information collection can take place during a pilot study.
In other cases, the prior information is collected during a pre-trial period, in which information is collected on the same subjects that will later participate in the full experiment.
For such information collected during a pre-trial period, the exact protocols for using such information should be determined in advance to ensure reproducible results.
For example, if results from the pre-trial period are used to eliminate some units from the study, the criteria for elimination should be predetermined and those units should not even have observations collected during the main study.

In this section, we discuss using controls as diagnostic tools to assess the quality of an experiment.
Another strategy is to incorporate these diagnostic statistics into inference procedures.
They could be used directly in an adjustment procedure, or indirectly in a sensitivity analysis.
For example, \citet{Tchetgen13}, \citet{Sofer2016}, and \citet{Miao18} use negative controls to adjust for confounding induced by unobserved covariates.

\subsection{Identifying unwanted factors}

Was our experiment properly designed, so that no unintended factors influence blood pressure?
In an ideal, well-controlled experiment, the only factor that should be allowed to influence blood pressure is the given treatment, caffeinated or decaffeinated coffee.
In a randomized experiment, a factor is an intervention manipulated by the researcher that has a possible influence on the outcome of interest.
Here, we use unintended factor to refer to an experimental condition which has a possible influence on the outcome, but which is not intentionally manipulated by the researcher.

A null treatment-control is one tool to help us detect whether a poorly-controlled experiment occurred.
Imagine that people have their blood pressure measured, are given a treatment, and then wait for 20 minutes before having their blood pressure measured again.
While waiting, some people sit in a waiting room with gardening magazines to pass the time, while others sit in a waiting room where the news is showing live updates of election results.
Some people in the local news waiting room may experience stress, which may result in an increase in their blood pressure.
In this scenario, blood pressure is being influenced both by the factor of interest, caffeine, and an unwanted factor, the stress of the election.
A null treatment-control could help the researcher assess whether unwanted factors may be influencing the outcome.

In order to implement a null treatment-control, we have three intervention groups: our standard groups receiving the active treatment and control treatment, and a third group which receives the null treatment-control.
The outcomes for all intervention groups will have unwanted variation due to the waiting room differences.
We can potentially detect this unwanted variation using the null treatment-control group.
Even for subjects who receive no intervention, we expect to observe some change in their short-term blood pressure due to measurement error.
In a well-controlled setting, these changes should be centered around zero.
Thus, we can calculate the average change in blood pressure given no intervention, i.e., the average of the primary outcome under the null treatment-control, as a diagnostic statistic.
If the average change is statistically not different from zero, we might have more confidence that no outside interventions occurred, even though we are aware that the null hypothesis can never be accepted.
On the other hand, in the waiting room setting, we might see that this diagnostic statistic is substantially positive due to some patients experiencing stress.
This result would encourage us to examine why there is an unexpected difference.

Is detecting this experimental mishap important?
If the waiting room assignment is random, then the effect of stress will be balanced on average across the active treatment and control treatment.
The resulting estimated treatment effect of interest would be unbiased, and thus would still be a valid estimate of the causal effect.
However, the estimate would potentially suffer from a higher variance, and thus a lower power to detect a treatment effect.
Unless we are in a setting with both a large sample size and a large signal, the higher variance is potentially a cause for concern.
Additionally, if for some reason the waiting room assignment were correlated with treatment assignment, then the treatment assignment mechanism could be confounded.
Perhaps the researchers used separate coffee machines in different rooms for the caffeinated coffee and decaffeinated coffee, and tended to place subjects in the waiting room closer to the coffee machine relevant to that subject.

In this contrived setting, such a correlation between treatment assignment and the outside factor seems unlikely, but such settings do occur in scientific studies.
For example, in genetics microarray studies, treatment assignment may be confounded because of its correlation with outside factors, which \citet{Gagnon12} term unwanted variation.
Consider a genetics study where the treatment of interest consists of a person's cancer status.
One lab conducts the microarray analysis on all of the samples from cancer patients, while a different lab handles all the healthy control samples. 
Different laboratory settings and experimental conditions have been shown to  influence the measured results of a microarray, leading to unwanted variation correlated with the treatment assignment.
This example is still somewhat unlikely; scientists now have a much better understanding of the variation in microarray results than when the technology was first being developed, and would try to avoid running samples in different settings if possible.
However, such an analysis plan could still feasibly be necessary for a large study conducted across multiple labs.
In this confounded setting, an unbiased estimate of the causal effect is no longer guaranteed.
Further diagnostic statistics can tell us if we are in this setting where treatment assignment is confounded; if we see that the average change in blood pressure across the whole sample is positive, we would next examine whether there is a difference between active treatment patients and control treatment patients.

A non-null treatment-control can also help us investigate our success in executing a well-controlled experiment.
The definition of a non-null treatment-control is that it has a known effect on blood pressure; in this case, we have chosen a hypertensive medication as a non-null treatment-control.
If we observe no increase in blood pressure when given hypertensive medication, we might question whether we are measuring change in blood pressure accurately.
For example, maybe different administrators are measuring the patients blood pressure for the `before' and `after' time points in the experiment, and one or more of the administrators are measuring the blood pressure improperly, perhaps differentially across groups.

A third diagnostic tool we propose for detecting unintended factors is a null outcome-control.
Flexibility is a null outcome-control, which means we have prior knowledge that it is not affected by drinking a cup of coffee.
Imagine we see that some people do show a change in flexibility after drinking coffee.
Upon further investigation, we discover that for some subjects, there was a mistake in carrying out the experiment; instead of leaving people for 20 minutes, they were left for 4 hours.
The long time period left people fatigued, decreasing their flexibility.
After 4 hours, any potential effects of caffeine on blood pressure will have dissipated, so we would not be able to detect any effect of caffeine.
As with the previous issue of the stressful waiting room, if this timing miscalculation is balanced across treatment and control, it would not bias our estimate of the treatment effect.
However, the poor execution of the experiment could reduce our power to detect any causal effect.

\subsection{Quality control in settings with substantial measurement error}
\label{sec:diag_error}

We have proposed using controls as diagnostic tools to detect flaws and ensure quality control in an experiment.
We now discuss possible specific decision rules for determining that the results are flawed due to substantial measurement error, possibly leading to the rejection of an experiment.

First, assume we have no measurement error in our primary outcome.
Then, given our null treatment-control of no intervention, the primary outcome of change blood pressure should be strictly zero for all subjects: $Y_i(w_{nt}) = 0$ for all $i$.
However, in the blood pressure example, and in many other settings, we do have measurement error.
In this case, the observed outcome is the latent potential outcome with some noise, e.g., noise is additive: $Y_i^{obs} \mathds{1}(W_i = w_{nt}) = \mathds{1}(W_i = w_{nt}) Y_i(w_{nt}) + \epsilon_i$.
Thus, if $w_{nt}$ is a valid null treatment-control, we would have a zero outcome $Y_i(w_{nt}) = 0$ as assumed, but we could still see the observed outcome being nonzero, i.e.,  $\mid Y_i^{obs} \mathds{1}(W_i = w_{nt}) \mid > 0$.
We can choose different diagnostic statistics, and corresponding decision rules based on these statistics, for determining whether to reject an experiment based on how strict we want our criteria to be.

Assuming a well-behaved and unimodal distribution of measurement error, we can choose simple diagnostic statistics.
One simple diagnostic statistic is the average change in blood pressure given the null treatment-control, $\overline{Y}_{w_{n_t}}^{p,obs} = {N_{w_{n_t}}}^{-1} \sum_{i:W_i = w_{n_t}} Y_i^{p,obs}$.
Based on our assumption that the null treatment-control results in a zero outcome, this number should be around zero.
To choose a decision rule based on this statistic, we can conduct a hypothesis test of $H_0: \overline{Y}^p(w_{nt}) = 0$, and set our decision rule to be that we determine the experiment is flawed if this hypothesis test is rejected.
An alternative diagnostic statistic is the mean of the absolute values of the observed changes: $\overline{Y}_{w_{nt}}^{abs, obs} ={N_{w_{n_t}}}^{-1} \sum_{i:W_i = w_{n_t}} \mid Y_i^{p,obs} \mid $.
We can also choose a threshold $A$, and then define the statistic of the number of units where the absolute value of their outcome exceeds the threshold $N_A = \sum_{i:W_i = w_{n_t}}  \mathds{1}(\mid Y_i^{obs} \mathds{1}(W_i = w_{nt}) \mid > A)$.
Then, our decision rule might be that we require $n_A = 0$ for the experiment to be accepted, or perhaps a less strict rule, such as $n_A < N/10$, that, e.g., no more than 10\% of the population sees a large fluctuation in change in blood pressure.

The decision rule to reject the results of an experiment should be determined in advance by subject-matter experts, based on their understanding of the setting, and based on how strict they want to be in catching potential experimental flaws.
If an experiment is easily repeatable, or if we are worried about how experimental quality can influence the results, we may choose a relatively strict decision rule.
On the other hand, if the potential flaws are not as consequential to the results of interest, we may instead choose a looser rule.
Researchers should consider all sources of variation when implementing such diagnostic tools.
For example, there could be heterogeneity in people's change in blood pressure in response to waiting in a doctor's office, which would result in variation in the observed responses unrelated to measurement error.
In this case, however, the diagnostic tool would function similarly as to a setting with measurement error.
The assumption of the null treatment-control is not strictly met for all units, and a diagnostic tool would help the researcher detect the heterogeneity, no matter the source, and decide whether the experimental design needs to be adjusted or not.

Implementing decision rules as discussed here results in some practical implications that are beyond the scope of this paper.
Using multiple diagnostic tests could result in a setting where issues with multiple testing arises.
However, often the main concern with multiple testing is false positives.
False positives in control tests could cause a researcher to throw away a legitimate experiment, but would not result in the typical problem of reporting a spurious result.
Additionally, power is an important consideration in these tests.
For example, the hypothesis test for the null hypothesis $H_0: \overline{Y}^p(w_{nt}) = 0$ could fail to be rejected merely because it is underpowered even if there is a problematic degree of measurement error.
Unfortunately, smaller studies are both often more repeatable and are more likely to be underpowered.
Ideally, diagnostic tests would be incorporated into the procedure for choosing an appropriate sample size to ensure the tests are sufficiently powered.

\subsection{Determining optimal timing}

How long should I wait to measure blood pressure after giving people an intervention?
We are conducting the caffeine study because we have a hypothesis that caffeine may affect blood pressure.
However, what if we do not know how quickly these effects will be measurable?
In order to detect a caffeine effect, we must measure blood pressure during the time window in which a subject is responding to caffeine; if we choose a time interval that is either too long or too short, then we will be unable to measure the effect of caffeine, and we will find a zero treatment effect of caffeine regardless of whether it had any influence or not.

In this case, we can run a small pilot study to collect data to determine the best time window that will detect any caffeine effects for most people.
One pilot study would be to use a non-null treatment control, such as a hypertensive medication, as our intervention, and then measure blood pressure at many different points after the intervention.
Alternatively, the pilot study could use the primary treatment, caffeinated coffee, but we do not yet know whether or not caffeine has a causal effect on blood pressure.
Instead, using the non-null treatment control gives us an experimental design for which we are confident a change in blood pressure is induced.
The strategy of using the non-null treatment control operates under the assumption that the primary treatment, caffeine, and the non-null treatment control, the hypertensive medication, operate on similar time scales.
In the caffeine example, this is perhaps plausible given that they are both orally administered substances that then need to be metabolized into the bloodstream in order to experience physiological effects.

A different pilot study design would be to use a non-null outcome control.
We are confident a non-null outcome control, in this case reaction time, has a nonzero outcome given the primary treatment, caffeine.
Thus, this pilot study would involve giving subjects a cup of coffee, and then measuring their reaction time at different time intervals to see when caffeine has its peak effect.
This second strategy also assumes that the outcomes of reaction time and blood pressure respond on the same time scale.
In the caffeine example, this is a plausible assumption given that once caffeine enters the blood stream, it has systemic effects.

When choosing between possible pilot designs, such as whether to use a non-null treatment-control or a non-null outcome-control, a main concern of the researcher may be cost and feasibility.
In some settings we can imagine that some aspects of the experiment are particularly difficult or costly, leading us to minimize the number of times we give a particular intervention or measure a certain outcome.
For example, imagine that the primary treatment is costly, in which case the researcher may choose the first pilot design, involving a non-null treatment control.
A non-null treatment control is necessarily a well-understood intervention, so it may be cheaper or have a smaller risk to administer.
For example, in a clinical trial, the primary treatment may be a novel medication, while the non-null treatment-control may be the standard treatment.
Alternatively, imagine that the primary outcome is difficult to measure, such as an outcome that requires state-of-the-art imaging.
Then, the researcher may want to choose the second pilot design, choosing a non-null outcome control that is much easier to measure than the primary outcome.

\subsection{Identifying responders and non-responders}

In many studies, there is population heterogeneity in the unit-level causal effect: different subjects respond differently to the primary treatment.
One simple example of heterogeneity is when there are two groups: some units are responders, in that the primary treatment has a nonzero effect on their outcomes, and other units are nonresponders, in that the primary treatment has no or only a negligible effect on their outcomes.
Another use of controls is to determine which units are responders and nonresponders.
This step can be conducted during a pre-trial period, using the same subjects as the main study.

In the caffeine example, some people in our population may be caffeine nonresponders, which we define as people who show no observable physiological responses due to caffeine.
Then, if we see no blood pressure change in these people, it is not because their blood pressure in particular does not respond to caffeine; it is because they do not respond to caffeine as a whole.
Under this definition, we are making a somewhat strict, but simplifying, assumption: if someone does not demonstrate certain specific physiological responses due to caffeine, then they are a nonresponder to caffeine as a whole, in that we assume other physiological responses that we do not observe will also be zero.
Essentially, we are assuming that caffeine has consistent responses across different types responses, and that someone will not be a partial responder and demonstrate some physiological responses, but not others.
During the pre-trial period, we can use non-null contrast-controls, which are contrasts in outcomes known to be nonzero given the primary treatment, to test for a subject's responder status.
In the caffeine example, non-null contrast controls include reaction time and alertness.
For each unit, during the pre-trial period, we measure reaction time with and without caffeine and estimate the causal effect of caffeine on their reaction time.
Then, we repeat the process with other non-null contrast controls.
If a particular subject has no estimated caffeine effect for the non-null contrast controls, then they are a caffeine-nonresponder according to our definition. 

Depending on our estimand of interest, we may or may not want to include nonresponders in our final study.
On the one hand, if we are interested in a population-level estimand, we would want to include both nonresponders and responders.
For example, a public health researcher determining the efficacy of implementing a caffeine reduction policy would likely be interested in the knowing how much caffeine affects blood pressure on average in the population.
The true population includes both responders and nonresponders, and that should be mirrored in our estimated causal effect.
In this case, although we include everyone in the study, using a pre-trial period to estimate responder status may still be of interest to gather more detailed information.
We may concentrate on estimating the causal effects separately for responders only, or we may be interested in the question itself of how many responders are in the population.
For example, consider a large and expensive study, in which the intervention fails to show efficacy in the broader population as a whole.
Collecting responder and non-responder status is one way to still potentially gain valuable information. 
The estimate of the population effect is driven both by effect size of the intervention on responders, and by the population breakdown into responder and non-responder status.
By identifying responder and non-responder status, we can decompose these two components, providing useful information about whether future research is warranted.
To avoid unconscious or conscious p-hacking, we recommend that the plan for defining responders and non-responders and for analyzing sub-groups be determined in advance.

On the other hand, if we are interested in unit-level causal estimands, we may want to exclude nonresponders from the study.
For example, for a doctor giving advice to a specific patient with high blood pressure, they may want to know the worst-case scenario for the effect of caffeine on the patient's health.
A doctor would want to know the effect for responders, ignoring any deflation in the causal effect that might occur by including nonresponders in the experiment.
If we have a limited amount of resources and are restricted to a particular study size, limiting the study to responders only would increase the power to detect the effect of caffeine in responders, or the precision of the estimate.
These issues are related to the idea of the generalizability of a study \citep{Dahabreh2020}; sometimes we would like to extend results from the population of the study to a different population of interest.

\subsection{Identifying compliers and non-compliers}

A similar strategy of using a pre-trial period can be deployed to determine if people are compliers and noncompliers.
Compliance during the pre-trial period can be measured using a non-null outcome-control.
Electrolyte concentration is a non-null outcome-control, which means it is known to be affected by drinking a cup of coffee; the fluid in coffee changes electrolyte concentration in the blood.
Imagine we see that some units do not show a change in electrolyte concentration, which leads us to uncover that some units did not like the taste of coffee, and threw it away instead of consuming it.
We may decide to exclude these noncompliers from the final study.
In the caffeine study, perhaps noncompliance is unlikely, and may also not be particularly costly to the experiment.
However, in other settings, such as large social science experiments involving implementing a costly intervention, this pre-trial step could potentially save a large amount of investment, as the final study will be much more likely to have a high proportion of compliers.
If the complier average causal effect is the primary estimand of interest, the researcher can best allocate resources to estimate their causal effect of interest with high precision.

\subsection{Summary of practical advice}

We briefly summarize some practical advice for researchers from this section.
One theme is that experimental controls can potentially be useful tools in understanding different sources of variation in an experiment, including experimental conditions, measurement error, and subpopulations such as compliers or responders.
Some of these types of variation, in particular those that induce confounding between the treatment assignment and the potential outcomes, can result in biased estimates of the causal effect.
Many other types do not necessarily result in biased estimates, but still can result in lower power to detect causal effects.

Another theme is that we advocate for researchers to consider implementing either a pilot study or a pre-trial period in which detailed information on controls is collected.
During the pilot study or pre-trial period, researchers can collect information about a wide variety of different controls that capture different experimental conditions in order to better understand potential sources of variation and their magnitudes.
Though adding an additional step is potentially costly, the extra step could ultimately lower costs if flaws in the experimental design are detected in advance of a large study being conducted.

Finally, the variety of examples shows that experimental controls must be tailored to the setting at hand.
We do not necessarily advocate that researchers need to implement every type of control for every study.
Instead, experimental controls can be used to target areas of an experiment that are most likely to result in unwanted variation.
By performing well controlled studies, knowledge is more efficiently and systematically built upon prior studies, as all controls are defined by assumptions from previous work.
The use of experimental controls should ultimately raise the quality of the work and push the field forward as controls used in every new investigation should reaffirm prior knowledge while generating new findings.

\section*{Acknowledgements}

This work was supported by the Department of Defense (DoD) through the National Defense Science \& Engineering Graduate Fellowship (NDSEG) Program, by the John Harvard Distinguished Science Fellow Program within the FAS Division of Science of Harvard University, and by the Office of the Director, National Institutes of Health under Award Number DP5OD021412. The content is solely the responsibility of the authors and does not necessarily represent the official views of the National Institutes of Health.

We thank Stephane Shao, Alice Sommer, Terry Speed, and members of the Miratrix CARES lab for their insightful comments on the manuscript.
We also thank the Bind Lab, Gagnon-Bartsch Lab, and members of the Harvard statistics department for their feedback.

\bibliography{ref}

\begin{thebibliography}{40}
\providecommand{\natexlab}[1]{#1}
\providecommand{\url}[1]{\texttt{#1}}
\expandafter\ifx\csname urlstyle\endcsname\relax
  \providecommand{\doi}[1]{doi: #1}\else
  \providecommand{\doi}{doi: \begingroup \urlstyle{rm}\Url}\fi

\bibitem[Alban et~al.(2006)]{Alban2006}
R.~F. Alban et~al.
\newblock Obesity does not affect mortality after trauma.
\newblock \emph{The American Surgeon}, 72\penalty0 (10):\penalty0 966--969,
  2006.

\bibitem[Angrist and Krueger(1999)]{Angrist1999}
J.~D. Angrist and A.~B. Krueger.
\newblock Empirical strategies in labor economics.
\newblock In O.~Ashenfelter and D.~Card, editors, \emph{Handbook of Labor
  Economics, Volume 3, Part A}, pages 1277--1366. Elsevier Science, 1999.

\bibitem[Boring(1954)]{Boring1954}
E.~G. Boring.
\newblock The nature and history of experimental control.
\newblock \emph{The American Journal of Psychology}, 67\penalty0 (4):\penalty0
  573--589, 1954.

\bibitem[Dahabreh et~al.(2020)Dahabreh, Roberston, Steingrimsson, Stuart, and
  Hern\'{a}n]{Dahabreh2020}
I.~J. Dahabreh, S.~E. Roberston, J.~A. Steingrimsson, E.~A. Stuart, and M.~A.
  Hern\'{a}n.
\newblock Extending inferences from a randomized trial to a new target
  population.
\newblock \emph{Statistics in Medicine}, 39:\penalty0 1999--2014, 2020.

\bibitem[Eisenberg and Levanon(2013)]{Eisenberg2013}
E.~Eisenberg and E.~Y. Levanon.
\newblock Human housekeeping genes, revisited.
\newblock \emph{Trends in Genetics}, 29\penalty0 (10):\penalty0 569--574, 2013.

\bibitem[Feller et~al.(2016)Feller, Grindal, Miratrix, and Page]{Feller16}
A.~Feller, T.~Grindal, L.~Miratrix, and L.~C. Page.
\newblock Compared to what? {V}ariation in the impacts of early childhood
  education by alternative care type.
\newblock \emph{The Annals of Applied Statistics}, 10\penalty0 (3):\penalty0
  1245--1285, 2016.

\bibitem[Flegal et~al.(2013)]{Flegal2013}
K.~M. Flegal et~al.
\newblock Association of all-cause mortality with overweight and obesity using
  standard body mass index categories.
\newblock \emph{Journal of the American Medical Association}, 1:\penalty0
  71--82, 2013.

\bibitem[Gagnon-Bartsch and Speed(2012)]{Gagnon12}
J.~A. Gagnon-Bartsch and T.~P. Speed.
\newblock Using control genes to correct for unwanted variation in microarray
  data.
\newblock \emph{Biostatistics}, 13\penalty0 (3):\penalty0 539--552, 2012.

\bibitem[Glass(2014)]{Glass14}
D.~J. Glass.
\newblock \emph{Experimental Design for Biologists}.
\newblock Cold Spring Harbor Laboratory Press, Cold Spring Harbor, New York,
  2nd edition, 2014.

\bibitem[Goh et~al.(2017)Goh, Wang, and Wong]{Goh2017}
W.~W.~B. Goh, W.~Wang, and L.~Wong.
\newblock Why batch effects matter in omics data, and how to avoid them.
\newblock \emph{Trends in Biotechnology}, 36\penalty0 (6), 2017.

\bibitem[Hartman and Hidalgo(2018)]{Hartman2018}
E.~Hartman and F.~D. Hidalgo.
\newblock An equivalence approach to balance and placebo tests.
\newblock \emph{American Journal of Political Science}, 62\penalty0
  (4):\penalty0 1000--1013, 2018.

\bibitem[Holland(1986)]{Holland86}
P.~W. Holland.
\newblock Statistics and causal inference.
\newblock \emph{Journal of the American Statistical Association}, 81\penalty0
  (396):\penalty0 945--960, 1986.

\bibitem[Keele et~al.(2019)Keele, Zhao, Kelz, and Small]{Keele19}
L.~Keele, Q.~Zhao, R.~R. Kelz, and D.~Small.
\newblock Falsification tests for instrumental variable designs with an
  application to tendency to operate.
\newblock \emph{Medical Care}, 57\penalty0 (2), 2019.

\bibitem[Kirk(1982)]{Kirk1982}
R.~E. Kirk.
\newblock \emph{Experimental design: procedures for the behavioral sciences}.
\newblock Brooks/Cole Publishing Company, Moneterey, CA, 1982.

\bibitem[Lazar et~al.(2012)]{Lazor2012}
C.~Lazar et~al.
\newblock Batch effect removal methods for microarray gene expression data
  integration: a survey.
\newblock \emph{Briefings in Bioinformatics}, 14\penalty0 (4):\penalty0
  469--490, 2012.

\bibitem[Lin et~al.(2019)]{Lin2019}
Y.~Lin et~al.
\newblock Evaluating stably expressed genes in single cells.
\newblock \emph{GigaScience}, 8:\penalty0 1--10, 2019.

\bibitem[Lipsitch et~al.(2010)Lipsitch, {Tchetgen Tchetgen}, and
  Cohen]{Lipsitch10}
M.~Lipsitch, E.~{Tchetgen Tchetgen}, and T.~Cohen.
\newblock Negative controls: {A} tool for detecting confounding and bias in
  observational studies.
\newblock \emph{Epidemiology}, 21:\penalty0 383--388, 2010.

\bibitem[Ma et~al.(2019)]{Ma2019}
J.~Ma et~al.
\newblock Drinking cold water increases blood pressure in healthy young
  students.
\newblock \emph{Blood Pressure Monitoring}, 19:\penalty0 118--19, 2019.

\bibitem[Miao and {Tchetgen Tchetgen}(2018)]{Miao18}
W.~Miao and E.~{Tchetgen Tchetgen}.
\newblock A confounding bridge approach for double negative control inference
  on causal effects.
\newblock \emph{arXiv:1808.04945}, 2018.
\newblock {https://arxiv.org/abs/1808.04945}.

\bibitem[Miller(2013)]{Miller2013}
J.~Miller.
\newblock Harvard researchers challenge results of obesity analysis.
\newblock \emph{The Harvard Gazette}, 2013.
\newblock
  {https://news.harvard.edu/gazette/story/2013/02/weight-and-mortality/}.

\bibitem[Mozer et~al.(2018)Mozer, Kessels, and Rubin]{Mozer18}
R.~Mozer, R.~Kessels, and D.~B. Rubin.
\newblock Disentangling treatment and placebo effects in randomized experiments
  using principal stratification -- an introduction.
\newblock In \emph{Quantitative Psychology}, pages 11--23. Springer
  International Publishing, 2018.

\bibitem[Pine et~al.(2016)]{Pine2016}
P.~S. Pine et~al.
\newblock Evaluation of the external rna controls consortium (ercc) reference
  material using a modified latin square design.
\newblock \emph{BMC Biotechnology}, 16, 2016.

\bibitem[Richardson et~al.(2014)Richardson, Laurier, Schubauer-Berigan,
  {Tchetgen Tchetgen}, and Cole]{Richardson14}
D.~B. Richardson, D.~Laurier, M.~K. Schubauer-Berigan, E.~{Tchetgen Tchetgen},
  and S.~R. Cole.
\newblock Assessment and indirect adjustment for confounding by smoking in
  cohort studies using relative hazard models.
\newblock \emph{American Journal of Epidemiology}, 180\penalty0 (9):\penalty0
  933--940, 2014.

\bibitem[Rosenbaum(2002)]{Rosenbaum2002}
P.~R. Rosenbaum.
\newblock \emph{Observational Studies}.
\newblock Springer Science+Business Media, New York, NY, 2nd edition, 2002.

\bibitem[Rosenbaum(2010)]{Rosenbaum2010}
P.~R. Rosenbaum.
\newblock Evidence factors in observational studies.
\newblock \emph{Biometrika}, 97:\penalty0 333--345, 2010.

\bibitem[Rubin(1974)]{Rubin1974}
D.~B. Rubin.
\newblock Estimating causal effects of treatments in randomized and
  nonrandomized studies.
\newblock \emph{Journal of Education Psychology}, 66\penalty0 (5):\penalty0
  688--701, 1974.

\bibitem[Rubin(1980)]{Rubin1980}
D.~B. Rubin.
\newblock Discussion of `{R}andomization analysis of experimental data in the
  {F}isher randomization test' by {B}asu.
\newblock \emph{Journal of the Americna Statistical Association}, 75:\penalty0
  591--593, 1980.

\bibitem[Rubin(2008)]{Rubin2008b}
D.~B. Rubin.
\newblock For objective causal inference, design trumps analysis.
\newblock \emph{The Annals of Applied Statistics}, 2\penalty0 (3):\penalty0
  808--840, 2008.

\bibitem[Rubin(2020)]{Rubin2019}
D.~B. Rubin.
\newblock The practical importance of understanding placebo effects and their
  role when approving drugs and recommending doses for medical practice.
\newblock \emph{Behaviormetrika}, 47:\penalty0 5–18, 2020.

\bibitem[Santos et~al.(2014)]{Santos14}
V.~G. Santos et~al.
\newblock Caffeine reduces reaction time and improves performance in
  simulated-contest of {T}aekwondo.
\newblock \emph{Nutrients}, 6\penalty0 (2):\penalty0 637--649, 2014.

\bibitem[Schuemie et~al.(2014)Schuemie, Ryan, DuMouchel, Suchard, and
  Madigan]{Schuemie14}
M.~J. Schuemie, P.~B. Ryan, W.~DuMouchel, M.~A. Suchard, and D.~Madigan.
\newblock Interpreting observational studies: why empirical calibration is
  needed to correct p-values.
\newblock \emph{Statistics in Medicine}, 33:\penalty0 209--218, 2014.

\bibitem[Shi et~al.(2020)Shi, Miao, Nelson, and Tchetgen]{Shi2020}
X.~Shi, W.~Miao, J.~C. Nelson, and E.~J.~T. Tchetgen.
\newblock Multiply robust causal inference with double- negative control
  adjustment for categorical unmeasured confounding.
\newblock \emph{Journal of the Royal Statistical Society: {S}eries {B}},
  82\penalty0 (2):\penalty0 521--540, 2020.

\bibitem[Shu and Yi(2019)]{Shu2019}
D.~Shu and G.~Y. Yi.
\newblock Causal inference with measurement error in outcomes: {B}ias analysis
  and estimation methods.
\newblock \emph{Statistical Methods in Medical Research}, 28\penalty0
  (7):\penalty0 2049--2068, 2019.

\bibitem[Sofer et~al.(2016)Sofer, Richardson, Colicino, Schwartz, and
  Tchetgen]{Sofer2016}
T.~Sofer, D.~B. Richardson, E.~Colicino, J.~Schwartz, and E.~J.~T. Tchetgen.
\newblock On negative outcome control of unobserved confounding as a
  generalization of difference-in-differences.
\newblock \emph{Statistical Science}, 31:\penalty0 248--361, 2016.

\bibitem[Splawa-Neyman et~al.(1923/1990)Splawa-Neyman, Dabrowska, and
  Speed]{Neyman1923}
J.~Splawa-Neyman, D.~M. Dabrowska, and T.~P. Speed.
\newblock On the application of probability theory to agricultural experiments.
  {E}ssay on principles. {S}ection 9.
\newblock \emph{Statistical Science}, 5\penalty0 (4):\penalty0 465--472,
  1923/1990.

\bibitem[{Tchetgen Tchetgen}(2014)]{Tchetgen13}
E.~{Tchetgen Tchetgen}.
\newblock The control outcome calibration approach for causal inference with
  unobserved confounding.
\newblock \emph{American Journal of Epidemiology}, 179\penalty0 (5):\penalty0
  633--640, 2014.

\bibitem[{The Global BMI Mortality Collaboration}(2016)]{GlobalBMI2016}
{The Global BMI Mortality Collaboration}.
\newblock Body-mass index and all-cause mortality: individual- participant-data
  meta-analysis of 239 prospective studies in four continents.
\newblock \emph{The Lancet}, 388:\penalty0 776–86, 2016.

\bibitem[Vandenbroucke et~al.(2016)Vandenbroucke, Broadbent, and
  Pearce]{Vandenbroucke2016}
J.~P. Vandenbroucke, A.~Broadbent, and N.~Pearce.
\newblock Causality and causal inference in epidemiology: the need for a
  pluralistic approach.
\newblock \emph{International Journal of Epidemiology}, pages 1776--1786, 2016.

\bibitem[Yang et~al.(2014)Yang, Zubizarreta, Small, Lorch, and
  Rosenbaum]{Yang2014}
F.~Yang, J.~R. Zubizarreta, D.~S. Small, S.~Lorch, and P.~R. Rosenbaum.
\newblock Dissonant conclusions when testing the validity of an instrumental
  variable.
\newblock \emph{The American Statistician}, 68\penalty0 (4):\penalty0 253--263,
  2014.

\bibitem[Yang(2006)]{Yang2006}
I.~V. Yang.
\newblock Use of external controls in microarray experiments.
\newblock \emph{Methods in Enzymology}, 411, 2006.

\end{thebibliography}

\end{document}